\documentclass{edpa}
\usepackage{latexsym}
\usepackage{amssymb}
\usepackage{epsf}
\author{Bernard S. KAY}
\title{Application of linear hyperbolic PDE to linear
quantum fields in curved spacetimes: especially black holes, time
machines and a new semi-local vacuum concept}

\address{Department of Mathematics, University of York, YORK YO10 5DD,
England, UK}
\email{bsk2@york.ac.uk}

\abstract 
{Several situations of physical importance may be modelled by linear
quantum fields propagating in fixed spacetime-dependent classical
background fields.  For example, the quantum Dirac field in a strong
and/or time-dependent external electromagnetic field accounts for the
creation of electron-positron pairs out of the vacuum. Also, the theory
of linear quantum fields propagating on a given background curved
spacetime is the appropriate framework for the derivation of black-hole
evaporation (Hawking effect) and for studying the question whether or
not it is possible in principle to manufacture a time-machine.  It is a
well-established metatheorem that any question concerning such a linear
quantum field may be reduced to a definite question concerning the
corresponding classical field theory (i.e. linear hyperbolic PDE with
non-constant coefficients describing the background in question) --
albeit not necessarily a question which would have arisen naturally in a
purely classical context.  The focus in this talk will be on the
covariant Klein-Gordon equation in a fixed curved background, although
we shall draw on analogies with other background field problems and with
the time-dependent harmonic oscillator.  The aim is to give a 
sketch-impression of the whole subject of Quantum Field Theory in Curved
Spacetime, focussing on work with which the author has been personally
involved, and also to mention some ideas and work-in-progress by the
author and collaborators towards a new ``semi-local'' vacuum
construction for this subject.  A further aim is to introduce, and set
into context, some recent advances in our understanding of the general
structure of quantum fields in curved spacetimes which rely on classical
results from microlocal analysis.}

\http{http://www-users.york.ac.uk/~bsk2/}

\begin{document}
\maketitle

\begin{section}{Introduction}\label{intro}

The subject of Quantum Field Theory in Curved Spacetime has long
attracted the interest both of some theoretical physicists and of some
workers in the theory of partial differential equations.

To explain what this subject is about, let us start, for example, with 
the relativistic wave equation (Klein-Gordon equation)  $$\left
(\frac{\partial^2}{\partial t^2} -\frac{\partial^2}{\partial x^2}
-\frac{\partial^2}{\partial y^2} -\frac{\partial^2}{\partial z^2} +
m^2\right )\phi=0$$ appropriate to the description of a free scalar
quantum ``particle'' of mass $m$ in the flat four-dimensional spacetime
of special relativity.  Equivalently, in a more compact notation, one
writes  $$(\Box + m^2)\phi=0\eqno{(\hbox{KG0})}$$  where $\Box$ stands
for $\eta^{ab}\partial_a\partial_b$, and $\eta=\hbox{diag}(1,-1,-1,-1)$
is the usual Minkowskian metric on ${\mathbb{R}}^4$.  (Above, we have
adopted the convention that the speed of  light $c$ is equal to 1.) 

Bearing in mind that, when we pass from Special to General Relativity,
the flat spacetime of Minkowski gets replaced by a curved spacetime
$({\cal M},g)$ -- ${\cal M}$ a four-dimensional manifold, $g$ a general
pseudo-Riemannian metric of signature $(+,-,-,-)$ -- it is natural
to generalize $(1)$ to the covariant Klein-Gordon equation
$$(\Box_g + m^2)\phi=0\eqno{(\hbox{KG})}$$ 
where $\Box_g$ now denotes the natural covariant generalization
of $\Box$ to this curved spacetime, i.e. the Laplace-Beltrami operator
associated with the metric $g$:
$$\Box_g=g^{ab}\nabla_a\nabla_b$$
$$=|\hbox{det}g|^{-1/2}\partial_a(|\hbox{det}g|^{+1/2}g^{ab}\partial_b)$$
$$=g^{ab}\partial_a\partial_b+\hbox{lower order terms}.$$
(Above, $g$ with indices upstairs denotes the inverse matrix to $g$ with
indices downstairs).

KG0 is, of course, the prototype equation for quantum field theory in
flat spacetime. It, and its higher-spin counterparts, when coupled
together with suitable (non-linear) interaction terms and when 
understood, not as classical PDEs but rather in an appropriate
quantum-theoretic sense, is believed to describe the behaviour of
elementary particles to the extent that the effects of gravity may be
ignored.  In much the same way, KG, for a given fixed background
spacetime $({\cal M},g)$, and when interpreted in a suitable quantum
sense -- it will be one of our purposes, below, to assign a definite
mathematical meaning to the resulting ``quantized KG'' -- is believed to
be a prototype equation for the effect of a given strong external
gravitational field on elementary particles propagating in its vicinity.
If one's principle interest is in the new features of quantum field
theory due to the presence of a strong external gravitational field, it
presumably suffices to study this simple linear model.  Of course, to
study such a theory is a far less ambitious thing to do than to attempt
to formulate a full theory of quantum gravity. In a full theory of
quantum gravity, not only would there be interactions between the
various matter (i.e. non-gravity) fields but also the gravitational
field would itself be dynamical and presumably require a quantum
description. But one believes that a theory such as quantized KG should
have an interesting domain of validity as an approximate theory, and
hopes that by studying it in its domain of validity one might even find
some clues as to the nature of quantum gravity itself.

During the last 30 or so years, the resulting subject of Quantum Field
Theory in Curved Spacetime and in particular,  quantized KG, has turned
out to be more rich and interesting than one might have expected.  
Before turning to the question just what is meant, mathematically, by
``quantized KG'' let us mention some of the physical predictions
resulting from its study.  (Note that, of course, physicists did not
wait for the mathematical definition of the theory to be completely
clear before starting to make their calculations!)

\begin{itemize}

\item \emph{The Hawking Effect:} It is the framework within which
Hawking made his spectacular 1974 prediction \cite{HawkEvap} that mini
black holes will not in fact be black but rather \emph{hot} with a
temperature, given in the case of a spherically symmetric black hole of
mass $M$  (and using units where $c=G=k=\hbar=1$), by the formula
$T_{\hbox{Hawking}}=1/8\pi M$.

\item \emph{The Time-Machine Question:} It also leads to interesting
results which suggest that time-travel-to-the-past scenarios, which,
worryingly, are seemingly permitted by classical general relativity, are
prevented from actually occurring by quantum effects.

\end{itemize}

At a conceptual level, the attempt to formulate the general theory
has raised some important matters of principle, the resolution of which
has arguably led to a deeper understanding of quantum field theory in
general, even if one is mainly interested in the Minkowski space case. 
These matters of principle are connected with the following, somewhat
paradoxical-seeming circumstance.  On the one hand:

\begin{itemize}

\item \emph{Particle Creation:} The principal physical phenomenon
associated with quantum field theory in curved spacetime is the creation
of pairs of particles out of the vacuum. 

\smallskip

Yet, on the other hand,

\smallskip

\item \emph{The Problematic Nature of ``Particles'':} In a general
curved spacetime context, the very notion of ``particle'', becomes vague
and ambiguous.  Correspondingly, the familiar Minkowski-space notion of
a single preferred ``vacuum'' state has to be abandoned and replaced
instead with a preferred family of physically permissible states,
amongst which a principle of democracy prevails.

\end{itemize}

A suitable conceptual-mathematical framework which reflects this state
of affairs is now fairly well developed (although there are still also
some important gaps as I'll discuss at the end of the talk in Section
\ref{locvac}). The key idea is to have a notion for the field itself
which does not rely on any particular concept of ``particles'' or
``vacuum'', and to regard this abstract notion of ``field'' as the
fundamental entity. This is achieved by adopting the so called
\emph{algebraic approach to quantum field theory} (see \cite{HaagBook}
and Chapter 3 in \cite{KayWald}) which involves the theory of
$*$-algebras and their states and Hilbert-space representations. 
However, as long as one is concerned with a prototype equation which is
\emph{linear} such as $(2)$, the  questions which arise concerning the
quantum theory ultimately  reduce to  questions about the underlying
classical partial differential equation; the role of the $*$-algebras
etc. being to establish a dictionary telling us \emph{which} classical
question corresponds to which quantum question. It seems justified, in
fact, to say that there is a:

\begin{itemize}

\item {\bf Metatheorem}: \emph{Any question concerning the quantum field
theory based on KG may be translated into a definite question concerning
KG -- regarded as a classical PDE (hyperbolic, and with non-constant
coefficients) -- albeit not necessarily a question which would have
arisen naturally in a purely classical context.}

\end{itemize}

Some of the questions which arise in this way have posed interesting
challenges and have led to the import into the subject of a wide variety
of techniques from the theory of linear PDE.  Especially, the Princeton
1992 PhD thesis of Marek Radzikowski \cite{RadThesis, RadI, RadII}  (see
Section \ref{alg} below) opened the way to solving some previously
unsolved problems in the theory with help of techniques and results from
\emph{microlocal analysis} \cite{DuiHorI, DuiHorII, HorBook}.  This
sparked off a revolution in the subject which is still at a very active
stage. See e.g. the recent papers of Fredenhagen, Koehler, Brunetti,
Verch, Junker, Sahlmann and Fewster \cite{BrunFredKoh, JunkRev,
JunkTalk,  VerchWave, BrunFred, SahlVerch, Fewster}).  Radzikowski's
work may be seen as having picked up where Duistermaat and H\"ormander
left off in 1972 when they discussed distinguished parametrices and the
Feynman Propagator \cite{DuiHorII} for KG and it should be mentioned
that Arthur Wightman  played an important role in influencing both the
earlier (cf. the last paragraph in the preface to \cite{DuiHorII}) and
(as Radzikowski's PhD supervisor) the later work and in keeping the
flame alive during the intervening twenty years.

The aim of the remainder of my talk will be to amplify on some of the
main physical ideas and results just mentioned and to state some
mathematical results related to them.  In particular, I shall aim to
provide some helpful physical background for workers in PDE wishing to
delve into the recent literature which applies microlocal
analysis.\footnote{For further general background reading on quantum
field theory in curved spacetime, see the textbooks \cite{BirDavBook,
WaldBook}.  Especially \cite{WaldBook} is close in spirit to the present
account.} I should emphasize that, aside from restricting my attention
to the simple model equation, KG, I shall also mainly limit my
discussion to work with which I have been personally involved.  
Nevertheless, I hope that what I have to say will provide a useful
entr\'ee not only to this work but also to
other recent work including, in addition to the papers using microlocal
analysis mentioned above, the recent mathematical work by several
authors on the Hawking effect (see papers cited in Section \ref{Hawk})
as well as the recent papers on spin-$\frac{1}{2}$ fields (which make
use of Dencker's work \cite{Dencker} on \emph{polarisation sets}) by
Kratzert \cite{Kratzert} and by Hollands \cite{Hollands}, and also the
papers on quantum (and the, related, classical) scattering theory by
Bachelot, Nicolas and Melnyk \cite{BachDirac, Nicolas, Melnyk} for
spin-$\frac{1}{2}$ fields and by Bachelot \cite{BachMaxwell} for
spin-$1$ fields on black holes.

Another purpose of this talk will be to set into context, and give
a brief account of, some recent work by myself and collaborators
towards a new ``semi-local vacuum'' concept.

In order to talk about so many things in such a short time, I shall make
liberal use of short-cuts, simple examples, and analogies.  

\end{section}

\begin{section}{Particle production: the harmonic oscillator analogy}

In many respects, the quantum theory of our Klein-Gordon equation, KG,
in an external gravitational field is analogous to the Klein-Gordon --
or Dirac -- equation in an external electromagnetic field.   In the
latter case, if the electromagnetic field is sufficiently rapidly
varying in time, it can cause the creation of electron-positron pairs
out of the vacuum.

\def\epsfsize#1#2{0.8#1}
\epsfclipon
\epsffile{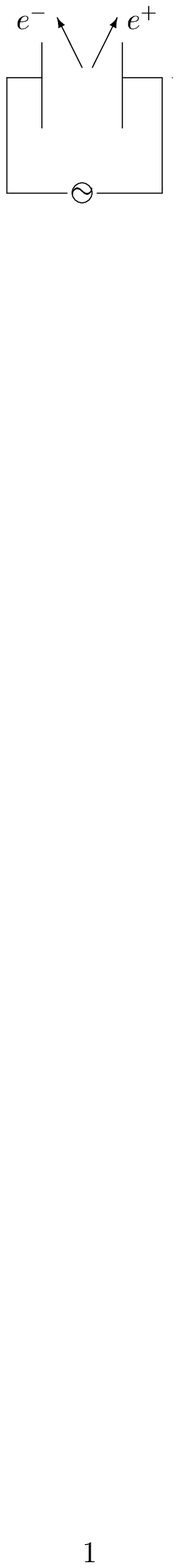}

We wish to explain how this pair creation comes about by  referring to
an even simpler analogy: the quantum harmonic oscillator with a
time-dependent frequency term.  As is familiar to everyone who has taken
an elementary course on Quantum Mechanics, this may be described by the
time-dependent Hamiltonian:

$$\hat H(t)=\frac{1}{2}\hat p^2 + \frac{1}{2}\omega^2(t)\hat x^2$$

Here $\hat x$ is to be thought of as analogous, in our electron-positron
example, to the field strength of the quantized Dirac field, and, in the
analogy with KG, to the $\phi$ field when it is suitably ``quantized'';
$\hat p$ is the quantum conjugate variable to $\hat x$, satisfying (we
shall take $\hbar$ to equal 1) $[\hat x, \hat p]=i$; and the term
$\omega^2(t)$ is analogous, in our electron-positron example, to the
field strength of the (classical) background electromagnetic field, and
in the analogy with KG, to the, generically, non-constant coefficients
in KG when it is written out in some coordinate system.

Suppose, for example, that $\omega^2(t)$ takes some constant value
$\omega_0^2$ before some ``initial'' time $T_1$,
returns to that same constant value after some ``final'' 
time $T_2$, and varies smoothly in between.

\def\epsfsize#1#2{0.8#1}
\epsfclipon
\epsffile{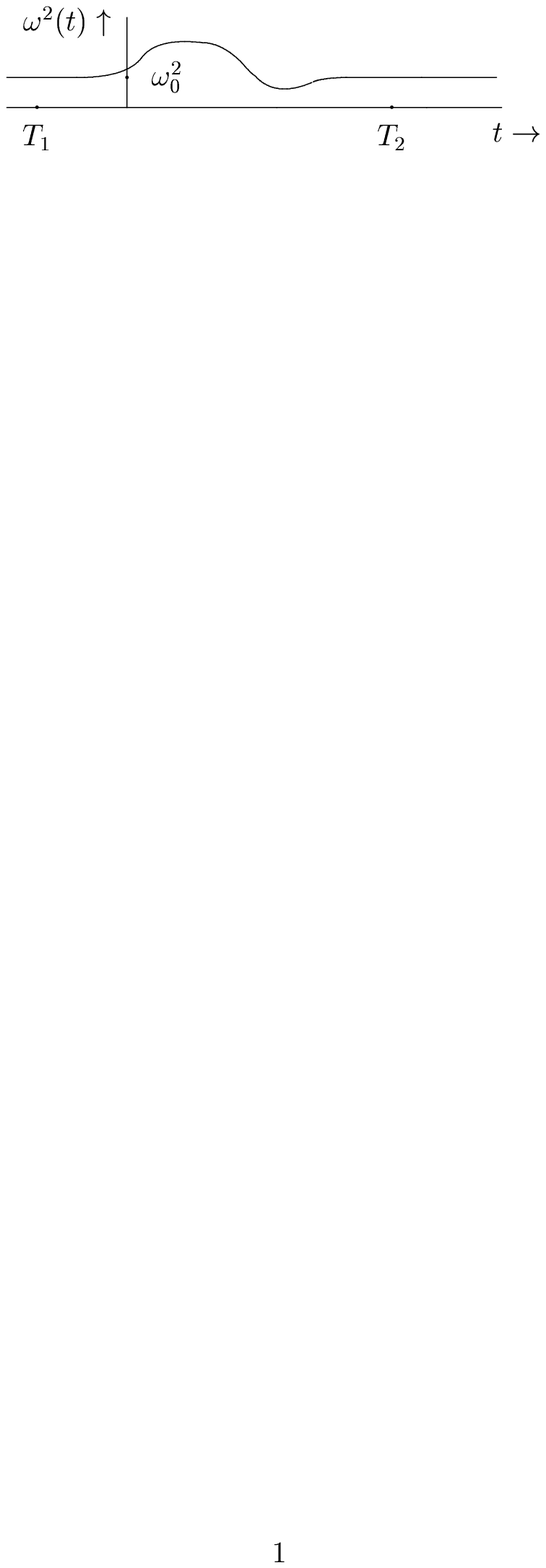}

Adopting the Schr\"odinger representation
$$\hat x \mapsto x, \quad\quad \hat p \mapsto -i\frac{\partial}{\partial
x}$$ $\hat H(t)$ maps to the differential operator
$$-\frac{1}{2}\frac{\partial^2}{\partial x^2}+\frac{1}{2}\omega^2(t)x^2$$
(on a suitable domain in $L^2_\mathbb{C}(\mathbb{R}^3)$).

The eigenfunctions of the constant $\hat H$ for the times earlier than
$T_1$ or later than $T_2$ are the usual harmonic oscillator wave
functions: $\psi_0(x)=c_0\exp(-\omega_0 x^2/2)$, $\psi_1=c_1
H_1(\omega_0^\frac{1}{2}x)\exp(-\omega_0 x^2/2)$,  $\psi_2=c_2
H_2(\omega_0^\frac{1}{2}x)\exp(-\omega_0 x^2/2)$, $\dots$ where $c_0,
c_1, c_2, \dots$ are normalization constants and $H_0, H_1, H_2, \dots$
the usual Hermite polynomials.  For the purposes of our analogy, these
should be thought of as ``vacuum state'',  ``one-particle state'',
``two-particle state'', etc.  

To understand the pair-creation phenomenon in the context of this simple
model, let us take the initial vacuum state $\psi_0$ and evolve it
according to the Schr\"odinger picture time-evolution for the
time-dependent Hamiltonian $\hat H(t)$ between times $T_1$ and $T_2$. 
In other words, let us consider the solution to the differential
equation (Schr\"odinger equation)
$$\left (-\frac{1}{2}\frac{\partial^2}{\partial
x^2}+\frac{1}{2}\omega^2(t)x^2\right )
\Psi(t,x)=i\frac{\partial\Psi(t,x)}{\partial t}$$
subject to the boundary condition $\Psi(T_1,x)=\psi_0(x)$.

By symmetry, the solution, $\Psi(T_2,x)$, to this problem will be
an even function of $x$, and
by completeness of the harmonic oscillator wavefunctions, it will
thus have an expansion:
$$\Psi(T_2,x)=a_0\psi_0(x)+a_2\psi_2(x)+a_4\psi_4(x)+\dots$$
(with $|a_0|^2+|a_2|^2+|a_4|^2+\dots=1$).
In general, $\Psi(T_2,x)$ will not itself be a multiple of $\psi_0(x)$.
In fact, as one may easily show, it will be a so-called
\emph{squeezed state} -- i.e. a Gaussian 
$\Psi(T_2,x)=c_\alpha\exp(-\alpha x^2/2)$ with a constant, $\alpha$,
in the exponent which, in general will differ from $\omega_0$. Thus
the coefficients $a_2$, $a_4$, $\dots$ will not vanish.  They
are to be interpreted as quantum amplitudes for the creation of
one, two, $\dots$ particle pairs.  

Thus we have fulfilled our promise of explaining how the phenomenon of
pair creation comes about.  But, staying with this simple analogy, we
can also get some insight into the sense (cf. Section \ref{intro}) in
which the concept of ``particle'' is vague and ambiguous:   Consider the
following sequence of possible modifications to the  problem just
discussed:  First, instead of the previous $\omega^2(t)$, consider an
$\omega^2(t)$ which takes on \emph{different} constant values, say
$\omega_{\hbox{in}}$ before $T_1$ and $\omega_{\hbox{out}}$ after $T_2$
and ask oneself the question: ``With respect to \emph{which} basis of
harmonic oscillator wave functions should we now expand $\Psi(T_2,x)$? 
Those where we substitute $\omega_0$ by $\omega_{\hbox{in}}$ or  those
where we substitute $\omega_0$ by  $\omega_{\hbox{out}}$?'' Next, return
to the original $\omega^2(t)$ but ask oneself how one could give a
particle interpretation to $\Psi(T,x)$ at a time $T$ lying between $t_1$
and $t_2$.  Finally, imagine an $\omega^2(t)$ which never settles down
to any constant value, either at early or late times!   As one considers
these situations in turn, it gets less and less clear how, and finally
impossible, to assign to any given state a definite
particle-interpretation.

\end{section}

\begin{section}{The field $*$-algebra and its states}\label{alg}

As we mentioned in the introduction, in view of the vagueness and 
ambiguity of the particle concept, one wants to have a mathematical
formulation which is not dependent on it.  This is achieved by adopting
the so-called algebraic approach to quantum field theory  (see
\cite{HaagBook} and Chapter 3 in \cite{KayWald}). To explain how this
works, we shall first say  what it amounts to for the familiar example
of the time-dependent harmonic oscillator just discussed:

First, one constructs the field $*$-algebra (with identity $I$) satisfied
by the (Hermitian) \emph{Heisenberg picture} $\hat x(t)$, at different
times.  This is determined once one specifies the commutator
$$[\hat x(t_1),\hat x(t_2)]=\hat x(t_1)\hat x(t_2)-\hat x(t_2)\hat x(t_1)$$
and one finds (e.g. one could show this with the elementary quantum
mechanical formalism discussed above)
$$[\hat x(t_1),\hat x(t_2)]=i\Delta(t_1,t_2)I$$
where $\Delta$ is the  difference of the classical advanced and retarded
Green functions for the \emph{classical} harmonic oscillator equation --
i.e. the unique bisolution  (i.e.
$(d^2/dt_1^2+\omega^2(t_1))\Delta(t_1,t_2)=0=
(d^2/dt_2^2+\omega^2(t_2))\Delta(t_1,t_2)$) to the \emph{classical}
harmonic oscillator equation which is antisymmetric ($\Delta(t_1,
t_2)=-\Delta(t_2,t_1)$) and satisfies ${\partial\Delta(t_1,t_2)/\partial
t_1}|_{t_2=t_1}=1$.  [This is the first manifestation of our
metatheorem.]

\bigskip

\noindent
\emph{Example:} If $\omega^2(t)=\omega_0^2 (=\hbox{constant})$, then
$$\Delta(t_1,t_2)=\frac{\sin\omega_0(t_1-t_2)}{\omega_0}.$$

\bigskip

Our various Gaussian states (now thought of as Heisenberg states,
unchanging in time) are now characterized by their symmetrized
two-point functions
$$G_\alpha(t_1, t_2)=\langle c_\alpha\exp(-\alpha x^2/2)|(\hat
x(t_1)\hat x(t_2)+\hat x(t_2) \hat x(t_1)) c_\alpha\exp(-\alpha
x^2/2)\rangle$$
which are to be thought of as constituting a democratic family; all
values of $\alpha$ being on an equal footing.  Again in accordance with
our metatheorem, this is a family of mathematical objects which may be
thought of as referring to the \emph{classical} theory, i.e. it consists
of symmetric bisolutions to the classical time-dependent harmonic
oscillator equation which satisfy an additional positivity requirement
(cf. Condition $(c)$ below) -- a set of objects related to the classical
differential equation, this time not one which would have arisen
naturally in a purely classical context.

\bigskip

\noindent
\emph{Example:} If $\omega^2(t)=\omega_0^2 (=\hbox{constant})$ and
also $\alpha=\omega_0$, then
$$G_{\omega_0}(t_1, t_2)=\frac{\cos\omega_0(t_1-t_2)}{\omega_0}.$$

\bigskip

All these mathematical structures generalize to our equation, KG,
provided we restrict our interest to the class of \emph{globally
hyperbolic} \cite{Leray} spacetimes.  We recall,  \cite{Geroch,
HawkEllBook, Dieckmann} that these consist of time-orientable spacetimes
which contain a Cauchy surface, and we shall assume below that a
particular choice of time-orientation has been made, so we can talk
about ``future'' and ``past'' in a meaningful way.  For this class of
spacetimes, there is  a natural analogue \cite{Lichn, ChoBru, Dimock},  
to the $\Delta$ we had for the harmonic oscillator,  which we shall call
the the \emph{Lichn\'erowicz commutator function} and also denote with
the symbol $\Delta$.  This is an antisymmetric distributional bisolution
to KG, and, with it, one immediately obtains a natural analogue for the
$*$-algebra of our harmonic oscillator example by quotienting the free
$*$-algebra with identity $I$ over $\mathbb{C}$ on abstract elements
$\hat\phi(f), f\in C_0^\infty({\cal M};\mathbb{R})$, by the commutation
relations 
$$[\hat\phi(f_1), \hat\phi(f_2)]=i\Delta(f_1, f_2)I$$ 
$\forall f_1, f_2\in C_0^\infty({\cal M};\mathbb{R})$, together with the
relations: $\hat\phi(a_1f_1+a_2f_2)=a_1\hat\phi(f_1)+a_2\hat\phi(f_2) \
\forall a_1, a_2 \in \mathbb{R} \ \forall f_1, f_2\in C_0^\infty({\cal
M};\mathbb{R})$ (i.e. linearity in test functions);
$\hat\phi(f)=\hat\phi(f)^* \ \forall f\in C_0^\infty({\cal
M};\mathbb{R})$ (i.e. Hermiticity); and $\hat\phi((\Box_g+m^2)f)=0 \ \forall
f\in C_0^\infty({\cal M};\mathbb{R})$ (i.e. the condition for $\hat\phi$ to
be a weak solution of KG).  The physical interpretation of the resulting
$\hat\phi(f)$ would then be as a ``smeared'' quantum field
``$\int_M\hat\phi(x)f(x)|\hbox{det}(g)|^\frac{1}{2}d^4x$'' for
test-function $f$.

Further, one may define natural analogues for the symmetrized two point
functions, $G_\alpha$, of our harmonic oscillator  Gaussian states to be
elements\footnote{The set of bidistributions, $G$ on ${\cal M}$,
satisfying Conditions $(a)$, $(b)$, and $(c)$ is in one-one
correspondence with the set of so-called \emph{quasifree states} (see
e.g. \cite{KayLib}).  This is only a small subclass of the set of all 
\emph{states} (in the ``algebraic sense'' \cite{HaagBook}, i.e.
positive, normalized, linear functionals on the field $*$-algebra).
However, for every non-quasi-free state, there will be a quasifree state
with the same two-point function, so the non-quasi-free states differ
only in their $n$-point functions for $n\ne 2$; cf. \cite{KayLib}. 
Moreover the quasi-free Hadamard states, i.e. those which in addition
satisfy Condition $(d)$, play an important role in that, as conjectured
in \cite{KayPad, KayComo} and proved in \cite{Verch}, they are
\emph{locally quasi-equivalent} and thus determine a unique local
\emph{folium} (see \cite{HaagBook}) of states and this is believed to be
the physically relevant local folium.} of the set of all bidistributions
on ${\cal M}$ --  i.e. of bilinear functionals   
$$G: C_0^\infty({\cal M};\mathbb{R})\times C_0^\infty({\cal
M};\mathbb{R})  \rightarrow \mathbb{R}$$  
satisfying the usual continuity properties -- such that
$\forall f_1, f_2, f \in C_0^\infty({\cal M};\mathbb{R})$,

\bigskip

\noindent
\emph{($a$) (symmetry)} $G(f_1, f_2)=G(f_2, f_1)$

\medskip

\noindent
\emph{($b$) (distributional bi-solution property)}
$G((\Box_g+m^2)f_1,f_2)=0= G(f_1,(\Box_g+m^2)f_2)$

\medskip

\noindent
\emph{($c$) (positivity)} $G(f,f)\ge 0$ \emph{and} $G(f_1,f_1)^\frac{1}{2}
G(f_2,f_2)^\frac{1}{2}\ge |\Delta(f_1, f_2)|$

\medskip

\noindent
\emph{($d$) (Hadamard condition)} ``$G(x_1,x_2)=-\frac{1}{2\pi^2}
\left (U(x_1,x_2)\hbox{P}\frac{1}{\sigma} +
V(x_1,x_2)\log|\sigma|+W(x_1,x_2)\right )$''

\bigskip

The last condition, the Hadamard condition, has no analogue for systems
with a finite number of degrees of freedom such as our 
one-degree-of-freedom harmonic oscillator model and is a restriction on
the nature of the (necessary) singularity in the ``unsmeared''  $G$
(elsewhere locally a smooth function) for pairs of points which are null
separated.  It is believed to be a physically necessary condition,
motivated in part by the equivalence principle, and in part by the
requirement that the state denoted by the $G$ in question should give a
finite expectation value for the quantum (renormalized) stress-energy
tensor (see especially \cite{Wald78}).  A na\"{\i}ve statement of the
condition is that the formula in quotes above should hold locally (and
in the sense of smooth functions for nearby pairs of non-null-separated
points) where $\sigma$, $U$, $V$, and $W$ are all defined on a
neighbourhood of the diagonal in ${\cal M}\times {\cal M}$; $\sigma$
denotes the squared geodesic interval between $x_1$ and $x_2$, $U$
(normalized so that $U(x,x)=1$) and $V$ are smooth two-point functions
which are determined by a standard procedure due to Hadamard ($U$ in
closed form and $V$, by certain \emph{Hadamard recursion relations}, as
a formal series in powers of $\sigma$ with coefficients which are smooth
two-point functions) in terms of the geometry alone, while $W$
(determined by another, standard, set of Hadamard recursion relations,
again as a formal series in powers of $\sigma$ with coefficients which
are smooth functions) also depends on, and characterises, the state in
question.  But to spell out a  mathematically meaningful statement of
the condition, more must be said to replace the formal power series by
genuine smooth functions, and also one apparently (but see Radzikowski's
local-to-global theorem below) needs to supplement this specification of
the singularity for nearby null separated points, by a statement to the
effect that $G$ is non-singular at all pairs of spacelike separated
points.  We refer to \cite{KayWald} for all the details.

What I wish to emphasize here is firstly that this condition has played
a very important role in most of the deeper mathematical results
concerning ``quantized KG'', secondly that the ``mathematically messy''
nature of the condition has been one of the major causes of technical
difficulties in establishing these results, and thirdly that, as I have
already mentioned in the introduction, an important breakthrough was
achieved in 1992 when Marek Radzikowski succeeded in replacing the
condition with a technically much cleaner statement  \cite{RadThesis,
RadI} couched in the language of  microlocal analysis, namely:

\bigskip

\noindent
\emph{($d'$) (Wave Front Set [or Microlocal] Spectrum Condition)}
$WF(G+i\Delta)=$

\smallskip

$\lbrace (x_1,p_1;x_2,p_2)\in T^*({\cal M}\times {\cal M}\setminus {\bf
0})\ | \ x_1$ and $x_2$ lie on a single null geodesic, $p_1$ is tangent
to that null geodesic and future pointing, and $p_2$ when parallel
transported along that null geodesic from $x_2$ to $x_1$  equals
$-p_1\rbrace$

\bigskip

\noindent
Here, we denote elements of the cotangent bundle of ${\cal M}$ by pairs
$(x,p)$, $x$ an element of ${\cal M}$, and $p$ a covector at $x$; ${\bf
0}$ denotes the zero section in $T^*({\cal M}\times {\cal M})$; and we
say that a covector is tangent to a curve at a point if the vector
obtained by ``raising an index'' with the metric is tangent to the curve
at that point.

Radzikowski in fact proved \cite{RadThesis, RadI} that, in the presence
of Conditions $(a)$, $(b)$ and $(c)$, Conditions $(d)$ and $(d')$
are equivalent.

\bigskip

\noindent
{\bf\large Examples}

\medskip

\noindent
\emph{Example 1}: $\Box\phi=0$ in Minkowski space $(\mathbb{R}^4,\eta)$ 
(i.e. KG0 with $m^2=0$):

\medskip

$$\Delta=\delta(\sigma)\varepsilon(t_1-t_2)$$
where $\sigma=\eta_{ab}(x_1^a-x_2^a)(x_1^b-x_2^b)$
and $\varepsilon(s)$ is the step function which is equal to $1$ if $s>0$,
and to $-1$ if $s<0$.

This example is of course so special that our principle of democracy
doesn't hold for it:  Exceptionally, there is a preferred symmetrized
two point function (i.e. that of the usual \emph{Minkowski vacuum state})
which we shall call $G_0$:
$$G_0=-\frac{1}{2\pi^2}\hbox{P}\frac{1}{\sigma}.$$
Equivalently, 
$$G_0+i\Delta=-\frac{1}{2\pi^2}\frac{1}{\sigma-2i\epsilon(t_1-t_2)-
\epsilon^2}$$
where we use the usual informal ``epsilon notation'' to denote a
distribution which arises as the boundary value of an analytic function.

\bigskip

\noindent
\emph{Example 2}: (Strictly, ``Examples 2 and 3'' are not examples but
rather lower dimensional analogues.  Especially, note that the
Hadamard condition $(d)$ needs to be adapted appropriately to
$1+1$ dimensions -- see \cite{KayBorCram}.) KG0 with $m^2=0$ in
1+1-dimensional Minkowski space $(\mathbb{R}^2,\eta)$.  This is of
course just the standard $1+1$-dimensional wave equation, which, with
the usual double-null coordinates $(U,V)$, may be written
$$\frac{\partial^2\phi}{\partial U\partial V}=0.$$

\medskip

$$\Delta=\frac{1}{2}\varepsilon(U_1-U_2)\varepsilon(V_1-V_2).$$

\def\epsfsize#1#2{0.8#1}
\epsfclipon
\epsffile{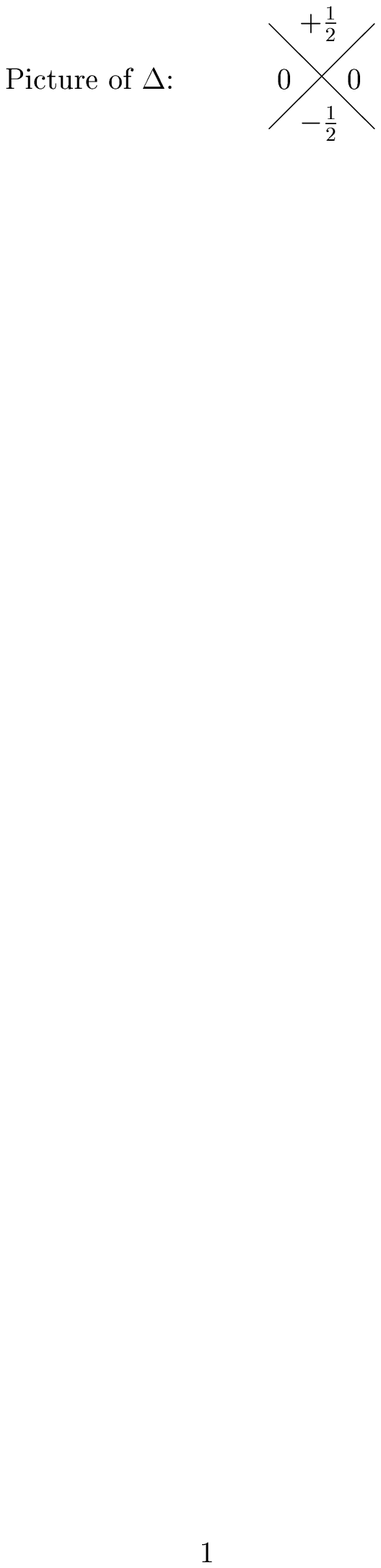}

Again, there is a preferred symmetrized two point function which we shall
also call $G_0$ and which is specified by
$$G_0+i\Delta=-\frac{1}{2\pi}\log[(U_1-U_2-i\epsilon)
(V_1-V_2-i\epsilon)].\eqno{(*)}$$
Actually, equation $(*)$ suffers from the well-known \cite{Wightman,
FulRui} ill-definedness of quantized massless fields in $1+1$
dimensions, and we should really say that $G_0+i\Delta$ only makes
unambiguous mathematical sense after it has been at least once
differentiated, and then that $(*)$ is just shorthand for the triplet of
equations:
$$\partial^2(G_0+i\Delta)/\partial U_1\partial
U_2=-(1/2\pi)(U_1-U_2-i\epsilon)^{-2}, \quad 
\partial^2(G_0+i\Delta)/\partial U_1\partial
V_2=0,$$
$$\partial^2(G_0+i\Delta)/\partial V_1\partial
V_2=-(1/2\pi)(V_1-V_2-i\epsilon)^{-2}.\eqno{(**)}$$

\bigskip

\noindent
\emph{Example 3}: KG with $m^2=0$ in a given $1+1$-dimensional (globally
hyperbolic) curved spacetime $({\cal M},g)$:  

\medskip

To analyze this example, we exploit the fact that one can always
(locally) find double-null coordinates, $(U,V)$ so that $g=C(U,V)dUdV$
for some $C^\infty$ function $C$ of two variables.  It is clear that
these coordinates are only fixed up to reparametrizations 
$$U\mapsto \hat U(U), \quad V\mapsto \hat V(V),$$
 -- below we shall refer to these as \emph{Virasoro reparametrizations}
because of the obvious resemblance to that concept from string theory --
and that, under these, $C(U,V)$ should be transformed to 
$$C(\hat U, \hat V)=\frac{dU}{d\hat U}\frac{dV}{d\hat V}C(U,V).$$
One easily sees that KG on $({\cal M},g)$ arises as the two-dimensional
wave equation of Example 2 with respect to any of these choices of
double-null coordinates. The appropriate choice of $\Delta$ is then
clearly locally identical with that of Example 2 for any choice of
$(U,V)$ coordinates; one easily sees that it is unchanged under Virasoro
reparametrizations.  On the other hand, equations $(*)$ (or if you
prefer $(**)$) above defining $G_0$ in Example 2 are \emph{not}
unchanged under Virasoro reparametrizations.  Because of this, one gets
lots of different symmetrized two-point functions, $G$, on $({\cal
M},g)$ each one  restricting, locally, to formula $(*)$ (or $(**)$) for
a different choice of $(U,V)$ coordinates. No one of these $G$ is to be
preferred over any other, in exemplification of our principle of
democracy.

\smallskip

\noindent
[\emph{end of Examples}]

\bigskip

The first application of the equivalence between Conditions
$(d)$ and $(d')$ was to the proof of the following theorem:

\medskip

\noindent
{\bf Local-to-Global Theorem (Radzikowski)} (a more general result is
proven in \cite{RadThesis, RadII}): \emph{Given a globally hyperbolic
spacetime $({\cal M},g)$ and letting $\Delta$ denote its Lichn\'erowicz
commutator function for KG, then if a bidistribution $G$ on ${\cal M}$
satisfies Conditions $(a)$, $(b)$, and $(c)$ globally on ${\cal M}$ and
also satisfies Condition $(d)$ separately on each element of an open
cover of ${\cal M}$, then it actually satisfies Condition $(d)$ globally
on all of ${\cal M}$.}

\medskip

This theorem confirmed the correctness of a conjecture which I had made
a few years earlier \cite{KayComo, GonKay}. Historically, it had been
the difficulty of settling this conjecture directly which had been the
stimulus for the work by Radzikowski which led to his discovery of the
Wave-Front Set spectrum condition $(d')$.

\end{section}  

\begin{section}{The Hawking effect}\label{Hawk}

As a preliminary, we first discuss the \emph{Unruh effect} which,
interestingly, was discovered simultaneously, in quite different 
contexts and with quite different motivations, by Unruh \cite{Unruh} and
by Bisognano and Wichmann \cite{BisWich}.  The setting for this is
quantum field theory in Minkowski space $(\mathbb{R}^4,\eta)$, and we
shall illustrate it with KG0 ($(\Box + m^2)\phi =0$).  Consider the
symmetrized two point  function, $G_0$, of the standard vacuum state,
but restrict attention to the right wedge $R=\lbrace (t,x,y,z)\in
\mathbb{R}^4 \ | \ x > |t|\rbrace$

\def\epsfsize#1#2{0.8#1}
\epsfclipon
\epsffile{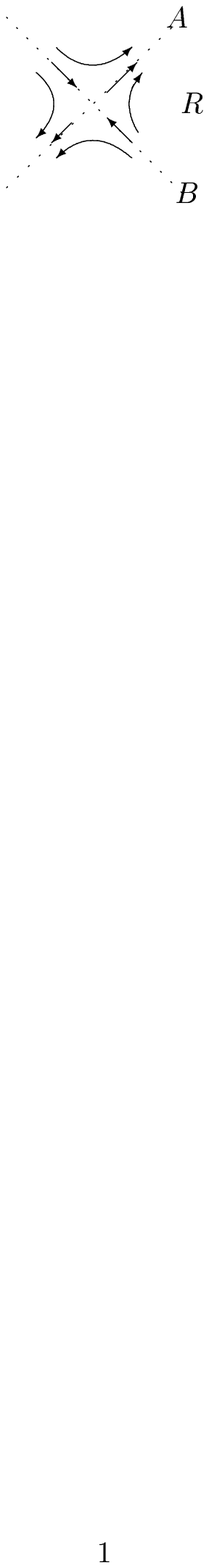}

and take ``time-evolution'' to be the one-parameter family of 
wedge-preserving Lorentz boosts.  Then the Unruh effect is the fact
that $G_0$ becomes the symmetrized two point function of a thermal
equilibrium state at ``temperature'' $1/2\pi$.

To illustrate how this comes about, take the $1+1$-dimensional massless
case (i.e. Example 2) and focus on the future boundary of the right
wedge, i.e. the right half of the null plane (now line!), $A$, i.e.
$\lbrace (t,x)\in \mathbb{R}^2 \ | \ t=x, t>0\rbrace$. In terms of the
$(U,V)$ coordinates of Example 2, this is $\lbrace (U,V)\in \mathbb{R}^2
\ | \ U>0, V=0\rbrace$ and Lorentz boosts act as dilations:
$U\rightarrow e^\tau U$ (where $\tau$ is the usual \emph{rapidity}).
Defining a new coordinate $u$ by $U=e^u$, these dilations become
$\tau$-translations and, in terms of them, one easily sees from Formula
$(**)$ in Example 2 above that

$$\frac{\partial^2(G_0+i\Delta)}{\partial u_1\partial u_2}=
-\frac{1}{2\pi}\left (\exp\left
(\frac{u_1-u_2}{2}\right)-\exp\left(\frac{u_2-u_1}{2}\right
)-i\epsilon\right )^{-2}$$
which is easily recognizable as the (twice differentiated) two point
function of a thermal state at ``temperature'' $1/2\pi$ restricted to
left-moving modes. (And a similar story of course holds for the past
boundary of the wedge/right-movers.)

There are many mathematical results related to the Hawking effect (see
e.g. \cite{HaagNS, DimKay,  FreHaag, BachColl, BachVac, BachHawk}). We
briefly discuss one of them.  Simplifying the wording slightly (the full
statement includes some further technicalities) a special case of it is:

\medskip

\noindent 
{\bf Theorem} (\cite{KayWald}, see also the generalization in
\cite{KayLib}): \emph{On the maximally extended Schwarzschild spacetime
of mass $M$, there is a unique Schwarzschild-isometry-invariant
bidistribution, $G$, satisfying Conditions $(a)$, $(b)$, $(c)$, $(d)$,
and moreover the corresponding quantum state, when restricted to the
exterior Schwarzschild region is a thermal state at the Hawking
temperature $1/8\pi M$.}

\medskip

We remark (cf. the classic paper of Rindler \cite{Rindler}) that to
visualize the Kruskal spacetime one can re-interpret the picture of
Minkowski space drawn above according to the substitutions:

\medskip

\noindent
Minkowski space $\rightarrow$ maximally extended Schwarzschild
spacetime

\smallskip

\noindent
null planes $A$ and $B$ $\rightarrow$ horizons

\smallskip

\noindent
right wedge, $R$ (each point a copy of $\mathbb{R}^2$)
$\rightarrow$ exterior Schwarzschild region  (each point a
copy of $\mathbb{S}^2$)

\smallskip

\noindent 
one-parameter family of wedge-preserving Lorentz boosts
$\rightarrow$ Schwarzschild isometries

\medskip

Moreover, the above theorem also respects this \emph{Rindler analogy}
in the sense that it remains true if we substitute $KG$ by $KG0$
and substitute the phrases to the right of the above arrows by the
phrases to the left, provided we also appropriately change the
Hawking temperature from $1/8\pi M$ to $1/2\pi$.

One of the key steps in the proof of this theorem, when adapted to the
latter Minkowski space version, is the demonstration that, for \emph{any}
bidistribution $G$ satisfying the conditions of the
theorem, 
$$\left.{\frac{\partial^2(G+i\Delta)}{\partial U_1\partial U_2}}
\right |_{V_1=V_2=0}
=-\frac{1}{4\pi}(U_1-U_2-i\epsilon)^{-2}\delta(y_1-y_2)
\delta(z_1-z_2).$$
Here we use coordinates $(U,V,y,z)$ where $U=t+x$ and $V=t-x$, so, 
geometrically, the restriction is to the null plane, $A$, which, back
in the Schwarzschild case, corresponds to a horizon.  

The uniqueness part of the theorem flows from the manifest uniqueness of
the right hand side of this equation.  The statement about thermality
flows from the observation that the right hand side of this equation is
identical (except for the delta function terms) with the two point
function for the Minkowski vacuum state of our Example 2, combined with
the remarks we made in the second paragraph of this section in
explanation of the Unruh effect for Example 2.

This demonstration was quite difficult and we refer to \cite{KayWald}
for the details.  The interested reader may find the following, much
easier, exercise a useful preliminary:

\medskip

\noindent
{\bf Exercise}: \emph{Show that the above formula holds in the case one
substitutes for $G+i\Delta$, the special value (i.e.  $G_0+i\Delta$ of 
Example 1 in the case $m^2=0$):}
$$G_0+i\Delta=-\frac{1}{2\pi^2}[(U_1-U_2-i\epsilon)(V_1-V_2-i\epsilon)-
(y_1-y_2)^2-(z_1-z_2)^2]^{-2}.$$

\end{section}

\begin{section}{Ruling out time-machines}

So far, we have stayed within the realm of globally hyperbolic
spacetimes.  But it is also interesting to ask to what extent the theory
might generalize to non-globally hyperbolic spacetimes.  Starting with
\cite{Thornetal}, one recently much discussed direction in which to
attempt generalization is to ask about spacetimes in which a
time-machine gets manufactured.  (See also \cite{KayFest} for other
non-globally hyperbolic spacetimes.)  Such a spacetime must, by
arguments due Hawking \cite{HawCPC}, schematically look like   

\def\epsfsize#1#2{0.8#1}
\epsfclipon
\epsffile{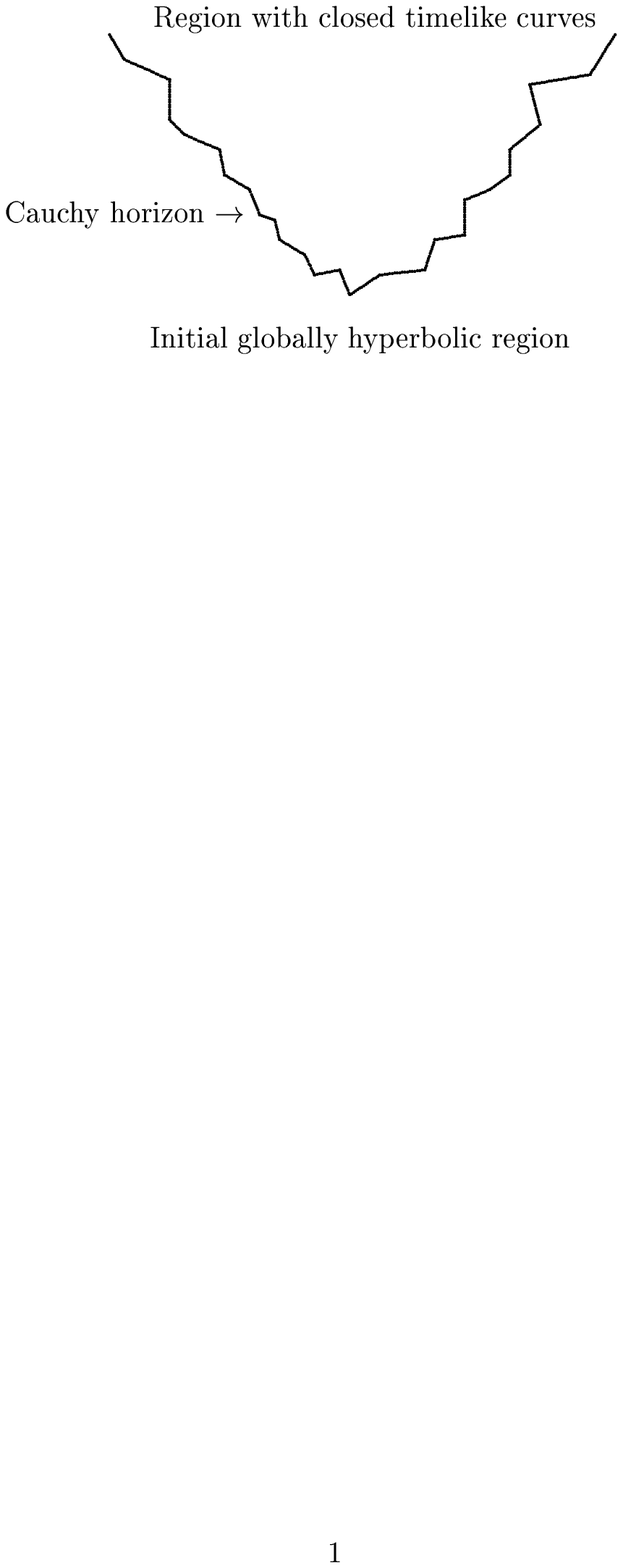}

The whole spacetime will be time-orientable and contain an initial
globally hyperbolic region, but there will also be a region with closed
timelike curves such that the two regions share, as their common
boundary, a \emph{compactly generated} (see \cite{HawCPC})  Cauchy
horizon.  (We recall that, in general, a Cauchy horizon is necessarily a
 -- not-necessarily smooth, see \cite{Chrusciel} -- null surface.) 
Within the framework described in the present talk, I, Radzikowski and
Wald (\cite{KayRadWald}, see also \cite{CraKay1, CraKay2}) obtained
several no-go theorems for a class of spacetimes which includes all
spacetimes which tend to suggest the conclusion that (at least
``semiclassically describable'' -- see \cite{Visser}) time machines are
not physically realizable. Special (actually weaker) cases of these
theorems are

\medskip

\noindent
{\bf Theorem 1} \cite{KayRadWald}: \emph{On such a spacetime, there
does not exist any antisymmetric distributional bisolution $\Delta$ such
that any neighbourhood of any point contains a globally hyperbolic
subneighbourhood on which $\Delta$ coincides with the intrinsic
Lichn\'erowicz commutator function of that subneighbourhood.}

\medskip

(In the terminology of \cite{KayFest}, Theorem 1 amounts to the statement
that this class of spacetimes is \emph{non-F-quantum compatible}.)

\medskip

\noindent
{\bf Theorem 2} \cite{KayRadWald}: \emph{On such a spacetime, there does
not exist any bidistribution, $G$, satisfying Conditions $(a)$ and $(b)$
and a weak local version of Condition $(d)$.}

\medskip

The proof of each of these theorems makes use of a geometrical lemma to
the effect that the Cauchy horizon of such a spacetime, $({\cal M},g)$,
must necessarily contain certain special points (called \emph{base
points} in \cite{KayRadWald}) with the property that every globally
hyperbolic neighbourhood ${\cal N}$ of a base point contains two other
points, $r$, and $s$, say, located inside the initial globally
hyperbolic region of $({\cal M},g)$ and such that $r$ and $s$ are
connected by a null geodesic in the total spacetime, but cannot be
connected by a causal curve lying within ${\cal N}$. In the case of each
theorem, this geometrical lemma is then easily combined with a suitable
microlocal \emph{propagation of singularities theorem} (\cite{HorBook}
Vol. IV), and the fact that on sufficiently small neighbourhoods of a
globally hyperbolic spacetime, $({\cal N},g)$,  the Lichn\'erowicz
commutator function of $({\cal N},g)$  (respectively, any
bidistribution, $G$, on $({\cal N},g)$ satisfying Conditions $(a)$ and
$(b)$ and the weak local version of $(d)$) is only singular for null
separated pairs of points, to obtain a contradiction.

\end{section}

\begin{section}{A new semi-local vacuum concept}\label{locvac}

The principle of democracy amongst our symmetrized two-point functions,
$G$, is very fine.  But, in practice, we \emph{can} tell the difference
between, say, an empty room and a room containing 10 tons of lead!

\def\epsfsize#1#2{0.8#1}
\epsfclipon
\epsffile{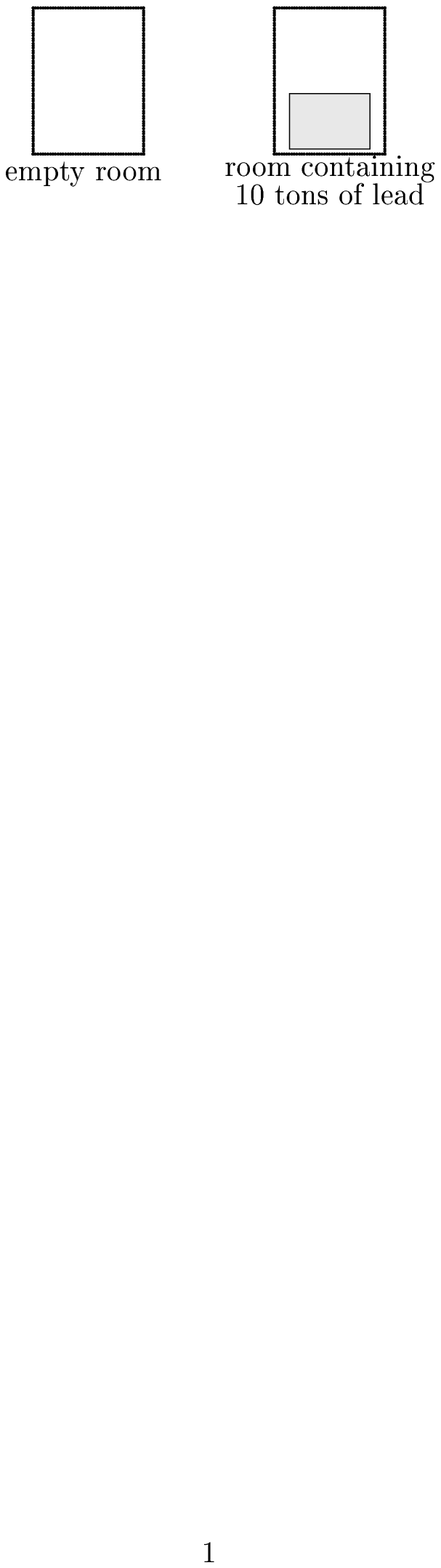}

As was recognized as early as 1976 by Hajicek \cite{Hajicek}, there
surely ought to be some way to reflect this in the theory of quantized
KG with an appropriate ``approximate local vacuum'' concept.  However,
up to now, as far as I am aware, no clean mathematical notion for such a
thing has ever been formulated.

I believe that what is lacking is something along the following lines:

\medskip

\noindent
{\bf Idea for a Definition}: Given a, say, globally hyperbolic spacetime
$({\cal M},g)$ and given any point $p\in {\cal M}$ then  \emph{a
symmetrized two point function, $G_p$, for a semi-local vacuum state
around $p$} is a  bidistribution, $G_p$, defined on a neighbourhood
${\cal N}_p$ of $p$ and satisfying conditions $(a)$, $(b)$, $(c)$ (with
$\Delta$ interpreted as the restriction of the Lichn\'erowicz commutator
function from ${\cal M}$ to ${\cal N}_p$) and $(d)$ on ${\cal N}_p$
which also satisfies the additional property that $G_p$ on ${\cal N}_p$
looks as closely as possible near $p$ to $G_0$ for $KG0$ on Minkowski
space.

\medskip

The problem is of course to decide on a suitable precise notion to
substitute for the phrase ``looks as closely as possible near $p$ to
$G_0$ for $KG0$ on Minkowski space''.  

In the case of the $1+1$ dimensional massless scalar field (i.e. our
Example 3) I have, with help from Andrew Borrott and Claes Cramer
\cite{KayBorCram}, arrived at the following notion which, it will be
argued in \cite{KayBorCram}, gives a very satisfactory solution to this
problem:  One defines $G_p$ amongst the class of $G$ described in
Example 3 by fixing the Virasoro invariance in $C$ with the
demands\footnote{I call such coordinates \emph{superspecial double-null
coordinates around $p$}.  These give a $C^\infty$ notion of ``local
vacuum''. One can also have a stronger notion of ``local vacuum'' by
taking what I call \emph{hyperspecial coordinates} i.e. by choosing
$(U,V)(p)=(0,0)$, $C(0,0)=1$, and demanding that, inside a suitable
neighbourhood, ${\cal N}_p$, of $p$, $C(U,0)$ and $C(0,V)$ take the
value $1$  throughout ${\cal N}_p$ -- i.e., thinking geometrically,
$C(r,p)$ and $C(p,s)$ take the value $1$ throughout ${\cal N}_p$ as $r$
and $s$ range over each of the two null geodesics which pass through
$p$.}: 
$$(U,V)(p)=(0,0),\quad C(0,0)=1,\quad \left.\frac{\partial^n C}{\partial
U^n}\right |_{(0,0)}=0=\left.\frac{\partial^n C}{\partial V^n}\right 
|_{(0,0)} \ \forall n\in{\mathbb{N}}.$$ 

I am presently working together with Stefan Hollands \cite{HolKay} on
some ideas towards a suitable counterpart to the above notion in the
case of massive fields and $1+3$ dimensions.

\end{section}


\end{document}